\documentclass[aps,prl,twocolumn,superscriptaddress]{revtex4}

\usepackage{graphicx} 
\usepackage{amssymb}

\usepackage{amsmath}

\usepackage{amsthm}

\usepackage{hyperref}

\usepackage{color,rotating}

\newcommand{\UV}{{\small UV}}
\newcommand{\IR}{{\small IR}}

\newcommand{\RG}{{\small RG}}

\newcommand{\QCD}{{\small QCD}}

\newcommand{\LHS}{{\small LHS}}

\newcommand{\STr}{\operatorname{Tr}}

\newcommand{\etaN}{\eta^{}_N}
\newcommand{\gN}{g^{}_N}

\newcommand{\fancysection}[1]{\vspace{.3\baselineskip}\addcontentsline{toc}{section}{#1}\textit{#1}~---}

\begin{document}

\title{Fixed points and infrared completion of quantum gravity}

\author{N. Christiansen}
\affiliation{Institut f\"ur Theoretische Physik, Universit\"at Heidelberg, Philosophenweg 16, 69120 Heidelberg, Germany}
\author{D. F. Litim}
\affiliation{Department of Physics and Astronomy, University of Sussex, Brighton, BN1 9QH, U.\,K.}                
\author{J. M. Pawlowski}
\affiliation{Institut f\"ur Theoretische Physik, Universit\"at Heidelberg, Philosophenweg 16, 69120 Heidelberg, Germany}
\affiliation{ExtreMe Matter Institute EMMI, GSI Helmholtzzentrum f\"ur
Schwerionenforschung mbH, Planckstr.\ 1, 64291 Darmstadt, Germany}
\author{A. Rodigast}
\affiliation{Institut f\"ur Theoretische Physik, Universit\"at Heidelberg, Philosophenweg 16, 69120 Heidelberg, Germany}

\begin{abstract}
  The phase diagram of four-dimensional Einstein-Hilbert gravity is
  studied using Wilson's renormalization group. Smooth trajectories
  connecting the ultraviolet fixed point at short distances with 
  attractive infrared fixed points at long distances are derived from
  the non-perturbative graviton propagator. Implications for the
  asymptotic safety conjecture and further results are discussed.
\end{abstract}

\maketitle

\fancysection{Introduction.} An important challenge in theoretical
physics relates to the quantum nature of gravity and the
incorporation of the gravitational force into the successful Standard
Model of Particle Physics. A promising avenue is provided through
Weinberg's asymptotic safety conjecture, according to which a quantum
theory of metric gravity may very well exist on a fundamental level
\cite{Weinberg:1980gg}.  The scenario stipulates the existence of an
ultraviolet {\small (UV)} fixed point which renders the theory finite even
beyond the Planck scale, in a manner similar to the well-known
weakly-coupled \UV{} fixed point of quantum chromodynamics {\small (QCD)}.  In
gravity, and unlike \QCD{}, the fixed point is expected to be interacting
and its investigation requires non-perturbative methods. Substantial
support in favor of a gravitational \UV{} fixed point has been
accumulated in recent years based on continuum studies, see
e.g.~\cite{Niedermaier:2006wt,Litim:2008tt,
Codello:2008vh,Litim:2011cp,Reuter:2012id}
and references therein, lattice simulations
\cite{Hamber:2009mt,Ambjorn:2012jv}, and holography
\cite{Litim:2011qf}.

A method of choice in the study of gravity in the continuum is given
by Wilson's renormalization group, based on the infinitesimal
integrating-out of momentum degrees of freedom. It permits a
systematic analysis even of strongly correlated and strongly coupled
theories and offers the prospect for a deeper understanding of gravity
in its extremes of shortest and largest distances. Furthermore,
insights achieved for other strongly correlated systems such as
critical scalar theories and \QCD{}, reviewed in
\cite{Litim:2010tt,Pawlowski:2010ht}, can now be exploited for
gravity.

A useful laboratory for quantum gravity is given by the
Einstein-Hilbert approximation, which retains Newton's coupling $G_N$
and the cosmological constant $\Lambda$ as the relevant couplings. By
now, the study of gravitational fixed points within the
Einstein-Hilbert theory and the phase diagram has lead to a consistent
picture with a non-trivial short-distance fixed point in Newton's
coupling and the cosmological constant, see
\cite{Niedermaier:2006wt,Codello:2008vh,Litim:2011cp,Reuter:2012id}
and references therein. Infrared fixed points corresponding to a
vanishing or negative cosmological constant
\cite{Donkin:2012ud,Litim:2012vz} have been detected as well. For
positive cosmological constants, the \IR{} behavior is plagued by
additional divergences, though the conjectured existence of an \IR{}
fixed point \cite{Bonanno:2001hi} has received some attention recently
\cite{Donkin:2012ud,Litim:2012vz,Nagy:2012rn,
  Rechenberger:2012pm,ContrerasLitim}.

In this Letter, we derive the fixed points and the phase diagram of
gravity from the \RG{} flow for the non-perturbative graviton
propagator. Our derivation of the \RG{} flow differs from previous
studies in two main aspects: firstly, we evaluate the flow on flat
backgrounds, which permits a better control on how the propagating
degrees of freedom drive the \RG{} flow for the relevant couplings. It
also provides a consistency check for earlier studies based on
background field methods~\cite{Niedermaier:2006wt,Litim:2008tt,
  Codello:2008vh,Litim:2011cp,Reuter:2012id}, bi-metric formulations,
\cite{Manrique:2010am}, and geometric flows
\cite{Donkin:2012ud}. Secondly, we adopt a new bi-local projection
technique to identify the scale-dependence of Newton's coupling and
the cosmological constant, offering an improved resolution of the
relevant fluctuations in the \UV{} and \IR{} limits of the theory.  As
a result, we obtain the first global phase diagram with a stable
ultraviolet fixed point which connects smoothly with an attractive
infrared fixed point at positive cosmological constant.

\fancysection{Gravitational renormalization group.}  The present work
is done within the functional \RG{} approach which is set up below. It is then
utilized for the derivation of the \RG{} flow for Newton's coupling $G_N$ and
the cosmological constant $\Lambda$. Under the \RG{} momentum flow, the classical
couplings $G_N$ and $\Lambda$ turn into scale-dependent couplings,
whose values depend on the \RG{} scale parameter $k$. As a function of
the latter, the couplings interpolate between the short $(1/k\to 0)$
and long distance regimes $(k\to 0)$ of the theory. It is convenient
to introduce the scale dependent, dimensionless couplings $\gN \equiv
k^2 G^{}_N/Z_{N,k}$ and $\lambda \equiv \Lambda_k /k^2$, where $Z_{N,k}$
denotes the graviton wave-function renormalization.  In
terms of these, the corresponding \RG{} $\beta$ functions have the form
\begin{equation}
\label{beta}
\begin{aligned}
 \partial_t \gN = \left( 2 + \etaN\right) \gN \quad  
\text{and} \quad
\partial_t \lambda =  
-2\lambda + \eta_{\lambda} \, ,  
\end{aligned}
\end{equation}
with 
$t\equiv\mathrm{ln} \left(k/ k_0\right)$ denoting the logarithmic \RG{}
`time', and $k_0$ an arbitrary reference scale.  We also introduced
the anomalous dimension of the graviton $\etaN \equiv -\partial_t
Z_{N,k}/Z_{N,k}$ and the `anomalous dimension' of the cosmological
constant $\eta_{\lambda} \equiv (\partial_t \Lambda_k)/k^2$. From
\eqref{beta} one can see that the fixed point condition $(\partial_t
\gN,\partial_t \lambda)=(0,0)$ is satisfied if the canonical running
due to the mass dimension of $G_N$ and $\Lambda_k$ is exactly
counter-balanced by the running induced by quantum fluctuations. In
case of an \UV{} fixed point this scaling is approached in the limit $k
\rightarrow \infty$, thus leading to divergence-free couplings at
arbitrarily small distances.  In order to deduce expressions for
$\eta_N$ and $\eta_\lambda$ we consider the following gauge-fixed
effective action, 
\begin{align}
\Gamma_{k}[\bar{g};h,\bar{C},C]
& = \frac{Z_{N,k}}{16 \pi G_N} \int \sqrt{|g|} \hspace{5pt}
(R(g)-2\Lambda_{k}) \notag \\  
& \mathrel{\hphantom{=}}+ \frac{Z_{\alpha,k}}{2 \alpha}\int
\sqrt{|\bar{g}|} \hspace{5 pt}\bar{g}^{\mu \nu} F_{\mu}(\bar{g},h)
F_{\nu}(\bar{g},h) \notag\\ 
&\mathrel{\hphantom{=}} -\sqrt{2} \int \sqrt{|\bar{g}|} \hspace{5 pt}
\bar{C}_{\mu} M^{\mu}_{\hphantom{\mu}\nu}(\bar{g},h) C^{\nu} \,.  
\label{EHtrunc}
\end{align}
In \eqref{EHtrunc} $R(g)$ is the Ricci scalar, and the second and
third terms are the gauge fixing term and the ghost action. We use the
harmonic gauge condition
$F_{\mu}(\bar{g},h) = \sqrt{2}
\left(\delta^{\beta}_{\mu} \bar{\mathcal{D}}^{\alpha} -
  \tfrac{1}{2}\bar{g}^{\alpha \beta} \bar{\mathcal{D}}_{\mu} \right)
h_{\alpha \beta}\,,$ where $\bar{\mathcal{D}}$
is the covariant derivative with respect to the background
connection. The background metric $\bar{g}$ is necessary to construct
the gauge fixing and allows for a split $g= \bar{g} + h$ with a
fluctuating graviton field $h$.  The background metric comes with its own auxiliary
background diffeomorphisms. The effective action is invariant under
combined transformations for the background and the fluctuating field
\cite{Litim:1998nf,Litim:1998qi,Freire:2000bq,Litim:2002ce,
Folkerts:2011jz}. The
Faddeev-Popov operator reads explicitly 
$M^{\mu}_{\hphantom{\mu}\nu} =
\bar{g}^{\mu \alpha}\bar{\mathcal{D}}^{\beta} \left(g_{\alpha \nu}
  \mathcal{D}_{\beta} + g_{\beta \nu} \mathcal{D}_{\alpha} \right) -
\bar{g}^{\alpha \beta} \bar{\mathcal{D}}^{\mu} g_{\beta \nu}
\mathcal{D}_{\alpha}$.
The scale-dependence of $Z_{\alpha,k}$ is irrelevant in Landau-DeWitt
gauge, $\alpha =0$ \cite{ Litim:1998qi,Folkerts:2011jz,Donkin:2012ud}
which we adopt throughout.
\begin{figure}
\begin{align*}
\partial_t \left. \frac{\delta^2\Gamma_k[\bar{g};h]}{\delta h^2}
\right|_{h=0} &= -\frac{1}{2} \includegraphics[height=6ex]{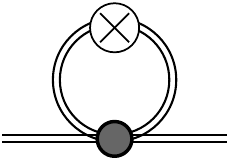} +
\raisebox{-2ex}{\includegraphics[height=6ex]{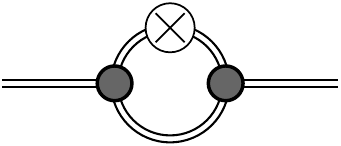}}\\
	      & \mathrel{\hphantom{=}} - 2
\,\raisebox{-2.2ex}{\includegraphics[height=6ex]{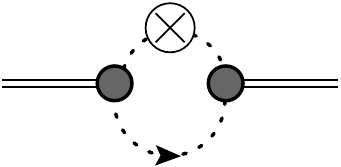}} \equiv \mathrm{Flow}^{(2)}
\end{align*}
\caption{Diagrammatic representation of the flow of the second order vertex
function. The dressed graviton propagator is represented by the double line, the
ghost  propagator by the dashed line, while a dressed vertex is denoted by a
dot and the regulator insertion by the crossed circle.}
\label{fig:flow_of_prop}
\end{figure}

We now turn to the scale-dependence of the effective action $\Gamma_k$
in \eqref{EHtrunc}. It is governed by \cite{Wetterich:1992yh}, 
\begin{equation}
\partial_t \Gamma_k[\bar g;\phi] = \frac{1}{2} \STr \,\frac{1}{\Gamma^{(2)}_k[\bar g;\phi] +
\mathcal{R}_k} \partial_t \mathcal{R}_k\,,
\label{floweq}\end{equation} 
where $\phi = (h,\bar{C},C)$.  The Wilsonian regulator $\mathcal{R}_k$
ensures finiteness of the equation, $(\Gamma^{(2)}_k +
\mathcal{R}_k)^{-1}$ is the full propagator and
$\Gamma^{(n)}_k\equiv\delta^n \Gamma_k / \delta \phi^n$.  The trace
over the operator product implies a sum over momenta, all internal
indices and fields with a minus sign for the ghost fields. We choose a
regulator of the form
\cite{Litim:2001up,Litim:2003vp,Pawlowski:2005xe}
\begin{equation}\label{opt}
\begin{aligned}
  \mathcal{R}_k \left(q^2\right) &= \Gamma^{(2)}_k\bigr|_{\lambda=0}\,
  r\left(q^2/k^2\right) \\
\qquad r(z)&=(\tfrac{1}{z}-1)\theta(1-z)
\end{aligned} 
\end{equation}
which allows for a largely analytical access. The \RG{} flow \eqref{floweq},
\eqref{opt} together with \eqref{EHtrunc} enables us to compute \eqref{beta}.

\fancysection{Propagator flow and flat backgrounds.}  The scale
dependence of the effective action $\Gamma_k$ in \eqref{EHtrunc} is
completely governed by that of $\Gamma^{(2)}_k$. It is only left to
specify our projection procedure for the flow $\partial_t\Gamma_k^{(2)}$:
Firstly we use flat backgrounds $\bar{g}_{\mu \nu}=\delta_{\mu
  \nu}$. This allows us to distinguish on the right hand side of the
flow between propagating modes and the non-dynamical background
\cite{Folkerts:2011jz}. This distinction is
crucial in Yang--Mills theory to obtain a confining
potential of the order-parameter \cite{Braun:2007bx}. Moreover, the
present setting allows us to study the dependence of $\Gamma^{(2)}_k$
on external momenta $p$. Following this route, we obtain the flow
equation by functional differentiation of \eqref{EHtrunc}. It has the
diagrammatic representation given in \autoref{fig:flow_of_prop}.  All
vertices and propagators are fully dressed.  The flow of the inverse
propagator contains the three- and the four-point function on the
RHS which follow by functional differentiation of
\eqref{EHtrunc}. The extensive algebra was carried out with the help
of \textsc{Form} and \textsc{xTensor} \cite{FORM,xTensor}. Note also
that the ghosts couple only linearly to the graviton in our
approximation.
\begin{figure}
\includegraphics[width=\columnwidth]{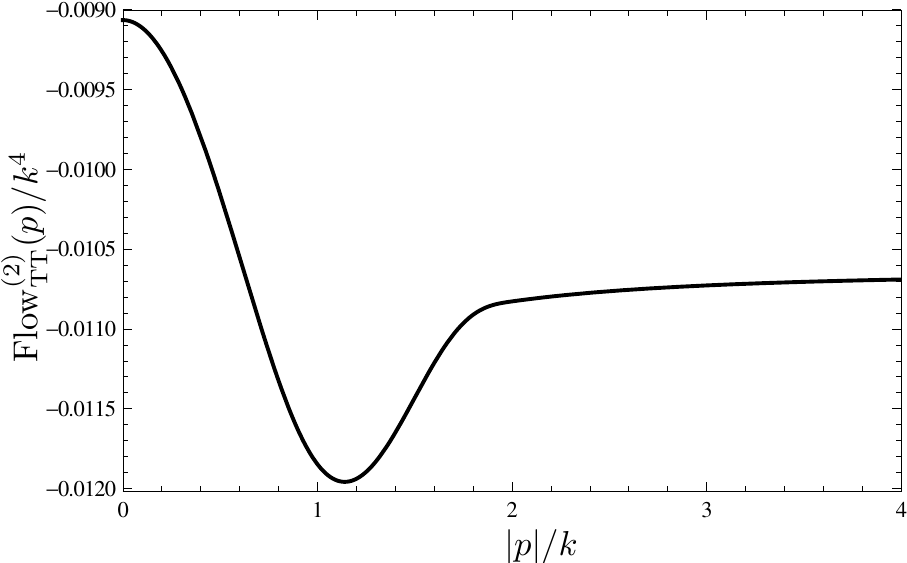}
\caption{Dependence of the flow on the external momentum $p$. In this
  plot for the parameter values $\lambda=0$ and $\etaN=0$.
} \label{fig:momdepflow}
\end{figure}

\fancysection{Einstein-Hilbert gravity.}  Next we turn to the
definition of the $\beta$-functions \eqref{beta} within the
Einstein-Hilbert approximation using an appropriate, bi-local,
projection in momentum space.
\begin{figure}
 \includegraphics[width=\columnwidth]{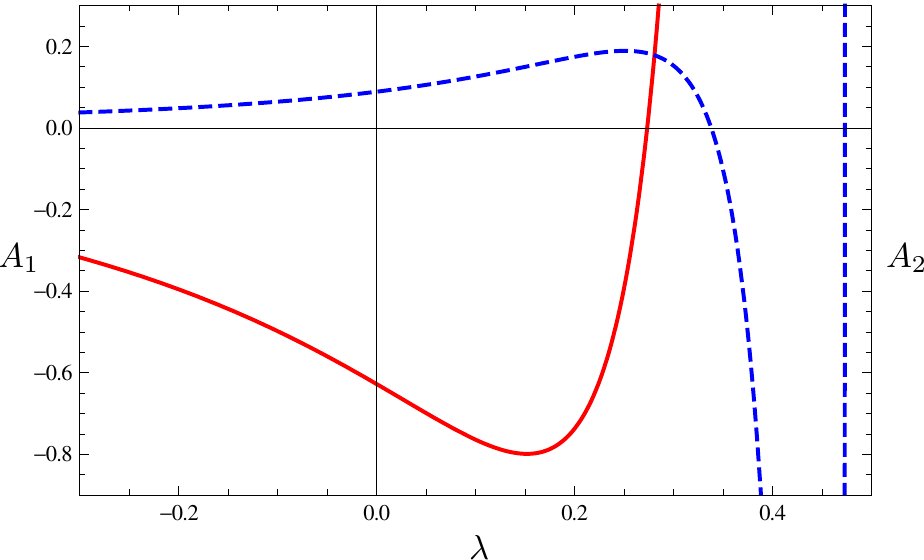}
 \caption{The functions $A_1(\lambda)$ (solid red line) and
   $A_2(\lambda)$ (dashed blue line) in the flow of $\etaN$
   \eqref{andim}. In the limit $\lambda\rightarrow1/2$ both functions
   behave as $(1-2 \lambda)^{-3}$. The function $A_2$ has a minimal
   value of $-13.6$ at $\lambda\approx0.46$.} \label{fig:Afcts}
\end{figure}
We recall that the relevant information is stored in the dynamical
spin two degrees of freedom of the graviton, which is obtained by
projecting the flow onto the transverse-traceless (TT) subspace of the
symmetric rank-four tensors.  Applying this TT projection to the
\LHS{} of the flow equation leads to
\begin{equation}
  \partial_t \Gamma_{k, \mathrm{TT}}^{(2)} = \frac{1}{32 \pi
    \, G_N}\left( p^2 \partial_t
    Z_{N,k} - 2  \partial_t \left(Z_{N,k}\Lambda_k\right)
  \right)\,.
\label{LHSflow}\end{equation}
From an analysis of the momentum structure one infers a quadratic
divergence of each diagram in \autoref{fig:flow_of_prop}, which we
have checked explicitly. Hence, one would expect a similar quadratic
divergence for large momenta of the right hand side of the TT-flow,
denoted with $\mathrm{Flow}_{\mathrm{TT}}^{(2)}(p)$. For large momenta
the flow can be calculated analytically and we find a cancellation of
all divergent terms, leading to the exact asymptotic behavior
\begin{equation}
\!\frac{\mathrm{Flow}_{\mathrm{TT}}^{(2)}(p)}{k^4}
\xrightarrow[p\rightarrow\infty]{} 
\frac{-20+42\lambda-48\lambda^2+(1+\lambda)\etaN}{192 \pi^2 (1-2\lambda)^2}.
\end{equation}
This general behavior of $\mathrm{Flow}_{\mathrm{TT}}^{(2)}(p)$ is
displayed in \autoref{fig:momdepflow} for $\etaN= 0$ and
$\lambda=0$. 

A similar cancellation has been observed in the context of the
Yang--Mills--gravity system where it leads to a vanishing
gravitational one-loop contribution to the running of the gauge
coupling \cite{Ebert:2007gf,Folkerts:2011jz}. To further exploit the
structure of the \RG{} flow for $\mathrm{Flow}_{\mathrm{TT}}^{(2)}(p)$
(see Fig.~2), we note that it displays a dip as a function of external
momenta $p$ at about $p \approx k$, with a curvature opposite to 
the one found at $p=0$.  Since fluctuations are integrated out at
momenta of the order of the \RG{} scale $k$, a consistent choice for the
evaluation of $\partial_t Z_{N,k}$ is given by the symmetric point
$p=k$, which we adopt to identify the \RG{} flow for Newton's
coupling. It also follows from \eqref{LHSflow} that the cosmological
constant appears as a graviton mass defined at vanishing
momenta. Therefore, the \RG{} flow for the vacuum energy is deduced from
$\partial_t (Z_{N,k}\Lambda_k)$ at $p=0$. Consequently, we are lead
to \eqref{beta} via a bi-local projection in momentum space, 
\begin{equation}
\begin{aligned}
    \partial_t \left(Z_{N,k}\Lambda_k\right) &= -16 \pi\, G_N\,
\mathrm{Flow}_{\mathrm{TT}}^{(2)}(p) \Bigr|_{p=0}\\
\partial_t Z_{N,k} &= \hspace{.8em} 16 \pi\, G_N\, \partial_p^2\,
\mathrm{Flow}_{\mathrm{TT}}^{(2)}(p)\Bigr|_{p=k} \,.
\end{aligned}
\label{Flow2}
\end{equation}
The virtue of our set-up is that the running of couplings is sensitive
to the global (momentum) behavior of the theory as encoded in its
two-point function. In addition, the flow of the cosmological constant
can be obtained completely analytically. Writing
$\eta_\lambda=\etaN\,\lambda+g\,A_3(\lambda,\eta)$, our \RG{} flow
\eqref{beta} takes the final form
\begin{equation}
\begin{aligned}
 A_3 &=
  \frac{1+4 (1-2\lambda)^2}{12 \pi
    (1-2 \lambda)^3} 
  - \etaN\frac{12 -45 \lambda -40 \lambda^2 }{180 
    \pi (1-2 \lambda)^3} +\frac{1}{\pi} \,,\\
\etaN &= \frac{A_1(\lambda) \gN }{1-A_2(\lambda)
\gN} - \gN \frac{237 \sqrt{3} - 160 \pi} {240 \pi^2} \,.
\end{aligned}
\label{andim}
\end{equation}
The last terms originate from the ghosts, and the functions
$A_1(\lambda)$ and $A_2(\lambda)$ are known numerically
and plotted in the relevant regime in \autoref{fig:Afcts}.

\begin{figure}
\includegraphics[width=\columnwidth]{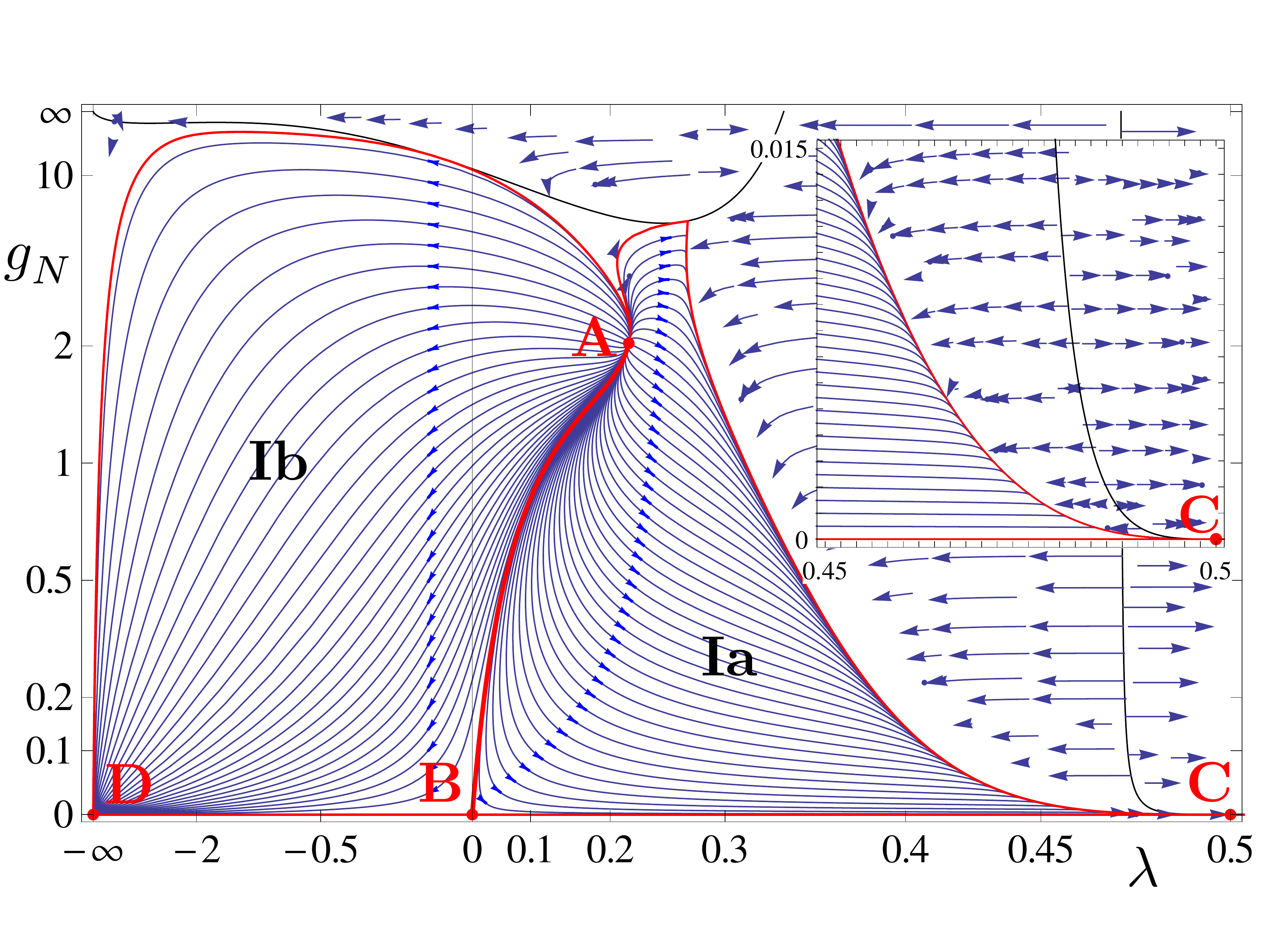}
\caption{Fixed points and phase diagram in the
  $(\gN,\lambda)$-plane. Arrows point from the \UV{} to the \IR{},
  red lines and points mark separatrices and fçixed points,
  black lines signal a divergence of the flow. The inset
  magnifies the vicinity of $C$.}
 \label{fig:phasediag}
\end{figure}

\fancysection{Fixed points and phase diagram.} 
\begin{table}
 \begin{tabular}{| c | c | c | c | c | c|} \hline
     & $g^\text{UV}_{N\,*}$ & $\lambda^\text{UV}_*$ & $\lambda^\text{UV}_*\times
g^\text{UV}_{N\,*}$ & $\theta_1$ & $\theta_2$
    \\ \hline
    no ghosts & 1.95 & 0.11 & 0.21 & $3.09 +2.00\, i$ & $3.09 -2.00\, i$\\
\hline
    with ghosts & 2.03 & 0.22 & 0.45 & $8.38$ & $2.60$ 
    \\ \hline
   \end{tabular}
\caption{Fixed point values ($g^\text{UV}_{N\,*}$,$\lambda^\text{UV}_*$), their
product, and the critical exponents  $\theta_{1,2}$ of
the
\UV{} fixed point without and with the ghost contributions.\label{table}}
\end{table}
The phase diagram of the \RG{} flows \eqref{beta} with \eqref{andim}
displays four connected fixed points $A$, $B$, $C$ and $D$ in the
physical regime, see \autoref{fig:phasediag}. The fixed point $A$ at
$(g^\text{UV}_{N\,*},\lambda^\text{UV}_*)\neq (0,0)$ denotes the
asymptotically safe \UV{} fixed point whose coordinates and scaling
exponents are given in Tab.~\ref{table}. It is characterised by two
\UV{} relevant real eigenvalues, which turn into a complex conjugate
pair in the absence of ghost field fluctuations. A complex conjugate
pair of eigenvalues, as found in many previous studies, can be lifted
via additional interactions e.g.~the inclusion of higher derivative
interactions. For our set-up, we conclude that the degeneracy in the
\UV{} scaling of the Ricci scalar and the vacuum energy is already
lifted by the ghost sector.  The fixed point $B$ at
$(g^\text{IR}_{N\,*},\lambda^\text{IR}_*)=(0,0)$ is the well-known
repulsive Gau\ss ian \IR{} fixed point. It corresponds to classical
gravity in the \IR{} with a vanishing cosmological constant. The fixed
point $D$ at $(g^\text{IR}_{N\,*},1/\lambda^\text{IR}_*)=(0,0^-)$ is
IR attractive in both couplings and governs infrared gravity with a
negative vacuum energy. In addition we find a non-trivial \IR{} fixed point $C$ at
\begin{equation}\label{IR}
(g^\text{IR}_{N\,*},\lambda^\text{IR}_*)=(0,1/2) \, ,
\end{equation}
that is connected to the \UV{} fixed point via smooth \RG{}
trajectories.  It is a distinctive property of the present approach
that it admits an infrared completion of quantum gravity for positive
cosmological constant. Naively, one would expect classical \RG-scaling
in the \IR{} fixed point regime \cite{Bonanno:2001hi}, for signatures
thereof see \cite{Litim:2003vp,ContrerasLitim}.  Here, we find an
asymptotic \IR{} behaviour with non-classical scaling exponents
$\Delta_g \approx 5.5$ and $\Delta_{(1-2 \lambda)} \approx 1.8$.  This
\IR{} behavior satisfies the exact scaling relation $\Delta_g = 3 \Delta_{(1-2
  \lambda)}$ within the given numerical accuracy. It implies an
asymptotic weakening of gravity, see also \cite{Donkin:2012ud}.  A
more detailed analysis will be presented in \cite{IRanalysis}. In our
set-up, the non-classical scaling of the propagator is a consequence
of strong \IR{} effects. It will thus be interesting to analyse the
impact of this scaling on the observable background Newton constant
and background cosmological constant, see \cite{Donkin:2012ud}.

It is important to stress that physical short-distance initial
conditions in the vicinity of the \UV{} fixed point $A$ leads to
trajectories in the regions Ia and Ib. In region Ia all trajectories
end into the new \IR{} fixed point $C$ with large anomalous dimensions
for the graviton propagator; in region Ib the trajectories end in the
\IR{} fixed point $D$ with classical scaling for the graviton
propagator. These two regions are
separated by the \IR{} instable separatrix leading to the Gau\ss ian
{\small FP} at $B$. In turn, all other trajectories terminate in
singularities and have no classical regime. Hence, the strong-gravity
region is shielded by the separatrices $AD$ ($g_N\!\leq\!33.5$)
and $AC$ ($g_N\!\leq\!5.4$) from regions which admit an extended
semi-classical regime. The fact that a strong-gravity regime is
shielded dynamically by the \RG{} flow is also present in previous
global \UV{}-\IR{} studies \cite{Donkin:2012ud,Litim:2012vz}, and thus
appears to be a generic feature of asymptotically safe gravity.

\fancysection{Summary.}  We have studied the phase diagram of
quantum gravity with a novel projection technique based on the
graviton two-point function. Our study also disentangles the role of
fluctuation and background fields, and confirms that gravity becomes
asymptotically safe. A further new result are global \RG{}
trajectories connecting the short distance fixed point with
weak-gravity long distance fixed points. Interestingly, the strong
gravity regime is dynamically shielded. 

\fancysection{Acknowledgements}  This work is supported by the Helmholtz Alliance
HA216/EMMI, and by the Science
and Technology Facilities Council (STFC) under grant number
ST/J000477/1.  

\end{document}